# On the apparently fixed dispersion of size distributions

Sascha Vongehr[a,*,**], Shaochun Tang[**], and Xiangkang Meng[**]

*Department of Physics, Nanjing University, Nanjing 210093, P.R. China
**National Laboratory of Solid State Microstructures, Department of Materials Science and Engineering, Nanjing University, Nanjing 210093, P.R. China

Probability density functions (PDF) of statistical distributions of cluster sizes $N$, where $N$ is the number of particles in the cluster, often seem to have less freedom than expected from considering the number of degrees of freedom at the clusters' source. The full width at half maximum appears to be comparable to the average $<N>$. Such a hidden symmetry is intriguing theoretically and practically impairs size selection towards narrow distributions. However, reviewing the example of Helium cluster beams demonstrates that the origin of the apparent fixing is the assumption that the distributions should be log-normal or exponential and the subsequent use of these functions to fit the data in $n = \ln(N)$ log-space. This demands more care when using parametric statistics. Alternatives to the traditionally employed fitting functions are discussed.

## 1. Introduction

Processes that randomize the results of previous random processes yet further give rise to typical statistical distributions. While decay or fractionation leads often to power laws[1,2] and exponential (EXP)[3,4] distributions, random grow processes like phase change aggregations give mostly rise to normal or lognormal (LN) results, be it in biology, economics or cluster physics[5,6,7]. Clusters' sizes can often be described by their mass or simply by the number $N$ of atoms or molecules inside the cluster. Sub- and supercritical expansions produce gas-condensation and fluid-fractionation clusters whose sizes $N$ seem distributed according to LN and EXP distributions, respectively. Considerations like entropy maximization given simple constraints support the assumption of these distributions. Especially in beams of clusters like for example beams of helium droplets He$_N$ (see reviews[8,9,10]), the clusters' sizes have a full width at half maximum (FWHM) proportional to the average size $\langle N \rangle$. Average and deviation are independent degrees of freedom (DOF) of a LN distribution. Nevertheless, for beams of He$_N$ from sub-critical expansions (i.e. gas condensations), LN size distributions and the restriction

$$\langle N \rangle / \text{FWHM} = 1.12 \pm 0.05 \qquad (1)$$

have been established[11]. Similar holds for the supercritical regime and the EXP function. Such results are however less than perfectly empirical because of the assumptions that went into the data collection. Besides assumptions regarding the physics of particular detection cross sections and collision cross sections, the main assumption of our concern in this work is the employment of rather restricted fitting functions, i.e. the use of parametric statistics. We show that the latter is responsible for the apparent (i.e. not actually empirical) fixing of the dispersion relation. The DOF, which are here the source stagnation pressure $P_0$ (20-100 Bar), temperature $T_0$ (10-30 K) and to some extend the nozzle size (diameter ~ 5 μm), seem reduced to the single parameter $\langle N \rangle$.

---

[a] Corresponding author

The fixing of dispersion relations is already interesting, but what makes this subject intriguing is the apparent numerical coincidence between different expansion regimes. It is very difficult to derive the size distribution within the complex physics of supersonic expansions, because the ideal gas law for example does not apply to cold helium. However, it is intriguing to notice that by cooling the cluster source down into the supercritical regime, the clusters become large fractionation clusters whose sizes seem

EXP distributed[12,13]. The EXP distribution is equi-dispersed, i.e. the standard deviation $\Delta_N$ exactly obeys

$$\Delta_N = \langle N \rangle \tag{2}$$

It is therefore often stated that "$\Delta_N \approx \langle N \rangle$" holds for both expansion regimes. This fact would be very surprising and thus triggered our investigations. The size distribution may only be determined by where the expansion trajectory through pressure $P_0$ versus temperature $T_0$ phase space intersects the bi-nodal and from which side it does so: When climbing the bi-nodal along the vapor side (i.e. cooling the nozzle), $\langle N \rangle$ increases and continues to increase when turning around at the critical point and descending along the liquid side of the coexistence curve. Although fragmentation is quite the opposite of growth, "$\Delta_N \approx \langle N \rangle$" suggest that cluster beam physics connects these regimes smoothly just by cooling the source, neglecting minor complications like that the supercritical expansion is somewhat bimodal due to re-condensing fragments. One may suspect that the proportionality between dispersion measures and averages has the same fundamental origin, or even that the LN distribution obtained when intersecting the bi-nodal from one side of the coexistence line can be derived from the EXP distribution present when intersecting from the other side. This makes the subject matter interesting for general cluster physics. A general coupling of the DOF in statistical growth processes would constitute a nuisance in need of exploration, because in nanotechnology, scaling up $\langle N \rangle$ and decreasing $\Delta_N$ simultaneously (size selection) are often both promised to be mere technicalities. This importance of so called size selection makes a fixed deviation very significant, which motivated us to investigate the dispersion relation

$$d_N := \langle N \rangle^2 / \Delta_N^2 \tag{3}$$

in detail theoretically and to review the usually employed methods that determine the parameters involved. In the following, we demystify the origin of the apparent fixing of the dispersion in unprecedented clarity and provide again an overdue cautionary notice about the dangers of parametric statistics, as done before in the difficult context of distribution mixing[14]. Because of the relevance to ongoing research and the ever important goal of size selection in cluster physics, we also provide alternative fitting functions and discuss their merits.

## 2. Introduction to Probability Density Functions

Our main subject is the origin of an apparent symmetry between different expansion regimes and the transformation from $N$ to log-space $n = \ln N$ will be part of the explanation. It is therefore worthwhile to pedagogically introduce the most convenient way to understand and manipulate probability density functions (PDF) of different distributions (like LN and EXP) and also their different expressions in $N$ and $n$ spaces together in one consistent notation.

<u>First we introduce the variables:</u> Consider a statistical variable $m$ in the real numbers with $m_{min} \leq m \leq m_{max}$ and average denoted as $\langle m \rangle$. The standard deviation $\Delta_m$ is defined via $\Delta_m^2 = \langle m^2 \rangle - \langle m \rangle^2$. Any linear transformation $m = an + b$ leaves its normalized variable $\tilde{n} = (m - \langle m \rangle)/\Delta_m$ untouched, meaning that $n = \Delta_n \tilde{n} + \langle n \rangle$ for all $n$ and the parameter

$$a = \Delta_m / \Delta_n \tag{4}$$

is a ratio of standard deviations. This can be also stated as $\langle \tilde{n} \rangle = 0$ and $\Delta_{\tilde{n}} = 1$. It helps to know that $n$ is $n = \ln N$ of the number of atoms $N$ in the cluster and that $m = \ln M$ is often the logarithm of the cluster's cross section. In that case, Eq. (4) is a fraction of spatial dimensions, so it will be insightful to conserve the difference between $m$ and $n$. The radii and cross sections of large clusters follow as $R = r_S N^{1/3}$ and $\sigma_{geo} = \pi R^2$ with the liquid Wigner-Seitz radius of $r_s$ = 2.221Å and 2.44Å for $^4$He and $^3$He respectively[15]. Nevertheless, these definitions are independent of the interpretations and with the further defining of $b = \ln B$, the equation $m = \ln(BN^{\Delta_m/\Delta_n})$ follows generally for any such $m$.

<u>Next we introduce the statistical distributions:</u> A convenient origin for a probability distribution is the cumulative probability $C$. Cumulative means that

$C = \int_0^{C_{(m)}} dC = \int_{m_{min}}^{m} (dC/dm)\, dm$, i.e. the infinitesimal probability of any $m$ is $dC$. The condition $C_{(m_{max})} = 1$ yields automatically normalized distributions for all $m = an + b$ if they are expressed as $C_{(\tilde{n})}$. The expectation value of any observable $\Psi$ is $\langle \Psi \rangle = \int_0^1 \Psi dC$, which, if boundaries are at infinity for example, equals $\langle \Psi \rangle = \int_{-\infty}^{+\infty} \Psi(dC/d\tilde{m})\, d\tilde{m}$. With $(d\tilde{m}/dm) = \Delta_m^{-1}$, the most likely value of $m$ (the "modal value" $\text{MODAL}_m$) is the position of the maximum where $(d^2C/dm^2) = 0$, as long as a maximum exists away from the boundaries at $m_{min}$ and $m_{max}$. $M = e^m$ implies $dM = M dm$ and the probability density functions for $M$ are therefore simply equal to the ones for $m$ yet divided by $M$:

$$\text{PDF:} = (dC/dM) = M^{-1}(dC/dm) \qquad (5)$$

## 3. The Size Distribution of Clusters from Sub-critical Expansions

When establishing the size distribution of sizes $N$ of clusters from sub-critical expansions, the data are usually fitted with a LN distribution[16,6], which can be for general $M$ easily gotten from the normal distribution (see appendix). The result is

$$\text{PDF}_{LN} = \frac{dC_{norm}}{dM} = \left\{ M\sqrt{2\pi}\Delta_m \exp\left[\frac{1}{2}\left(\frac{m-\langle m \rangle}{\Delta_m}\right)^2\right]\right\}^{-1} \quad (6)$$

and depends on two DOF, namely the fitting parameters $\langle m \rangle$ and $\Delta_m$ in log-space $m = \ln M$. In case of our cluster sizes $N$, the average $\langle N \rangle$ and the FWHM are then calculated (not measured) employing Eq. (11) etc. For Helium droplets $He_N$ it follows the well known proportionality[11] of Eq. (1). However, one should note something curious and important for the present work, namely that

$$d_M = \langle M \rangle^2 / \Delta_M^{\,2} = \left[e^{\Delta_m^2} - 1\right]^{-1} \quad (7)$$

, i.e. the dispersion of the LN is independent of $\langle m \rangle$ almost as if one DOF disappeared! This is however only one aspect of several that make the FWHM an especially misleading measure. The FWHM can be expressed via (see derivation in appendix)

$$\langle N \rangle / \text{FW}X\text{M} = \exp\left[3\Delta_n^2/2\right] / \left[\exp\left[\Delta_n\sqrt{-2\ln X}\right] - \exp\left[-\Delta_n\sqrt{-2\ln X}\right]\right] \quad (8)$$

Firstly, the choice of employing the FWHM, i.e. $X = ½$, determines the proportionality factor of $-2\ln X = \ln[4]$. There is nothing fundamental about the choice $X = ½$ and $X = 0.44$ would have lead to $\langle N \rangle = \text{FW}X\text{M}$ exactly. Secondly, the data derived fitting parameter $\langle n \rangle$ is actually not inside Eq. (8) at all. Thirdly, the proportionality varies weakly with $\Delta_n$. In fact, the surprising relation in Eq. (1) would be equally true after setting the original data for $\Delta_n$ to a constant 0.61 instead. All this is easily overlooked, and the reason is a combination of the following:

1) The $\langle n \rangle$-independence of the dispersion relation $d_N$ of LN distributions [from Eq. (7)] is hidden by the usage of $\langle n \rangle$ and $\Delta_n$ as fitting parameters and the subsequent transformation (the move from $n$- to $N$-space) into two variables, namely $\langle N \rangle$ and FWHM, that depend strongly on $\langle n \rangle$. 2) The FWHM naturally scales with $\langle N \rangle$, is insensitive to the large "foot" of the distribution, and is an especially bad measure of dispersion for parameters like particle number $N$ that have an absolute minimum.

It can be convenient and insightful to manipulate large quantities with absolute zero points, like entropy, in log-space. However, one should avoid transforming forth and back, especially when introducing assumptions due to models at different stages, because a slight variation in $n$ corresponds to a large change in $N$. In the sub-critical range $\langle n \rangle = 7 \pm 3$ of the helium experiments with continuous beams done to date holds

$$\Delta_n = 0.55 \pm 0.15 \tag{9}$$

However, $\Delta_n$ actually does depend on $\langle n \rangle$, for example[11]:

$$\langle n \rangle \approx 14.7 \Delta_n \tag{10}$$

This in turn makes the dispersion $d_N$ dependent on $\langle n \rangle$; i.e. it is not as independent as Eq. (7) and (1) suggest. Using the original data[11] results in Figure 1, a plot of the square of the often presented $\langle N \rangle / \text{FWHM}$ and of the dispersion relation $d_N$ [Eq. (3)], which one might erroneously assume to be quite similar or proportional to each other. Both are presented versus $\langle n \rangle$ along the horizontal axis. While the former measure stays surprisingly constant, the dispersion relation $d_N$ decreases strongly and illustrates the large experimental uncertainty better.

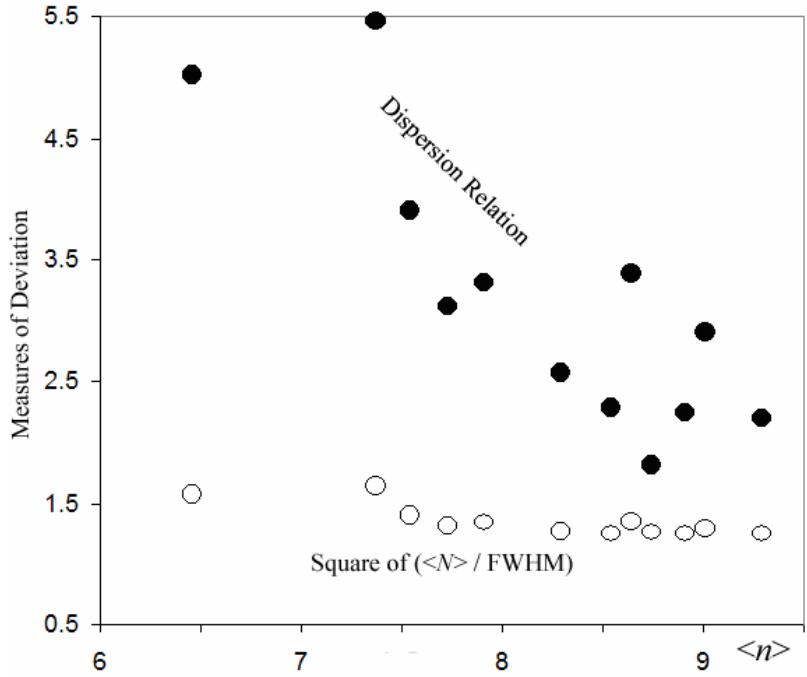

**Fig 1:** Two measures of the deviation of cluster sizes $N$ are plotted versus the mean *<n>* of $n = \ln(N)$: The square of an often presented measure, namely the average *<N>* divided by the FWHM, remains fairly constant (white dots) while the dispersion relation $d_N$ decreases strongly with *<n>* (black dots).

Since parametric statistics is a convenient way of analyzing data, we need to offer alternative fitting functions. It has been said that the strong influence of atomic evaporation on Helium cluster sizes makes it immediately doubtful that one should assume simple size distributions. However, one should not think in terms of simplicity. It is insufficient to merely select PDF with more DOF. A PDF of a different kind with as many DOF may work better. For smaller droplets, the low density of the droplets' surface increases the geometrical cross section much over the simple liquid drop model $\sigma_{l.d.}$. This suggests that cross sections may not be LN distributed, yet, given the accuracy of experiments, $N$, $\sigma$, both or even neither (in case of a certain correction[17] to the cross section) deviate from a LN distribution. As long as $\Delta_n \leq 0.6$ (compare Eq. (9)), the Inverse Gaussian (IG) distribution[18,19] traces the LN extremely well. If $M$ is the cross section, then $\Delta_m = (2/3)\Delta_n$ [Eq. (4)] is even smaller. Therefore, sub-critical cluster size distributions or their statistical ensemble of cross sections etc. could have been modeled with the IG's $\mathrm{PDF}_{(d_M, \langle M \rangle)} = \sqrt{\dfrac{d_M \langle M \rangle}{2\pi M^3}} e^{\frac{-d_M}{2M\langle M \rangle}(M-\langle M \rangle)^2}$ all along. This facilitates the Poisson mixture necessary to improve the prediction of cluster scattering and impurity pick-up statistics[14], which is also involved in the experiments that established the size distributions, because they do infer the size $N$ from the deflection of the cluster when hit by a probe beam of fast particles. The IG is useful for cluster physics because its Poisson mixture[20] is a closed expression; the LN's mixture is not. The LN and the IG are special cases of the Power IG[21,22] distribution, which has one DOF more, namely the power $p$ with $p = 0$ and $p = 1$ yielding the LN and IG respectively. The Power IG can fit the distribution of $M$ also in the supercritical regime. To employ a distribution with one more DOF seems complicated, but one may fix the DOF via a relation like Eq. (10). The Generalized IG distribution (GIG) includes many others as special cases ($\Gamma$, Hyperbolic, IG, Reciprocal IG (RIG), …) and also allows tractable Poisson mixture[23]. In any case, the assumption of an LN in order to fit the data is not without alternative and certainly not excluded by the data.

## 4. The Size Distribution of Clusters from Supercritical Expansions

In the supercritical regime, liquid helium fragments into droplets. At large $N$, the size distribution falls off exponentially[12]. The average $\bar{N}$ determined by fitting an EXP to large clusters $N > \langle N \rangle$ results in a higher outcome than averaging all clusters[13]:

$\bar{N} = (1.18 \pm 0.05)\langle N \rangle$ (experiments were done with $\langle N \rangle \approx 7*10^6$, i.e. $\langle n \rangle \approx 15.2$ if one assumes an EXP distribution). This disagreement is due to detection cross sections. Only the decay slope of the signal's logarithm for large $N$ above the average equals that of the original droplet distribution. Moreover, similarly large variations of 20% in calculated $\langle N \rangle$ at the same nozzle conditions argue for yet more unknown experimental artifacts. Considering that deflection experiments are moreover incorporating the assumption of a simple Poisson collision statistics, then all one can deduce in conclusion is that the original droplet size distribution is similar to a member of the exponential family.

Using a linear exponential for $N$ gives it a special status: If $N$ is EXP distributed, $M$ is not. This is different from the LN, where $M$ is automatically LN distributed if $N$ is. Therefore, one may use the gamma ($\Gamma$) distribution Eq. (12) instead. For $d_N = 1$ it yields the EXP distribution $\text{PDF}_{\text{EXP}} = e^{-N/\langle N \rangle}/\langle N \rangle$, but it should be noted that it always has the exponential fall off that is observed at high arguments $N$. Similar to what is encountered in case of the LN distribution above, the assumption of a simple, linear EXP distribution introduces the fixed dispersion $d_N = 1$. This fixing can be expressed in log-space (see appendix), where $\Delta_N = \langle N \rangle$ translates into $\Delta_n = \pi/\sqrt{6} \approx 1.28$. This is maybe the clearest expression of that the use of the EXP in order to fit data is not an innocent assumption, since $\Delta_n$ is completely fixed by it. One should recall that $\Delta_n$ is a data derived fitting parameter in all the research concerning the sub-critical regime and that it varies along with $\langle n \rangle$. The by investigations of the supercritical regime implied $\Delta_n = 1.28$ is purely due to the fitting procedure and it would be very surprising indeed if it were actually true even when generously allowing large errors due to measurement inaccuracy.

## 5. Conclusion

We analyzed the apparent fixing of the dispersion of cluster size distributions and the numerical coincidence of that fixing across different expansion regimes. The origin of this curiosity lies partly in the use of parametric statistics, i.e. the strong influence of the assumed probability distributions used to fit the data. Focusing on the example of Helium droplets, it turned out that the assumption of an LN or EXP distribution in order to fit the data in log-space $n = \ln N$ is the actual origin of the apparent symmetry between the expansion regimes.

The assumptions of the fitting functions are not innocent and we have shown that alternative distributions are not excluded by the data sets and their experimental uncertainties. The LN, EXP and also the simple Poisson distribution (via the modeling of cluster scattering) are always implicit in the data. These distributions dominate practically without alternatives in the Helium droplet community, but this is partially a historical fact and stabilized by that one cannot find any data that not already depends on the implicit use of the assumptions when fitting curves. Mathematical convenience is not a sufficient justification for neglecting to consider other distributions, because using an IG distribution in the sub-critical or Gamma functions in the supercritical regimes for instances, can be similarly or even more convenient, for example when the distribution has to be folded with formulas for detection efficiencies, beam depletion, impurity pick-up etc. We recommend these alternative distributions. As a further conclusion of more practical value to the experimentalist, we cannot support the suspicion of a hidden symmetry that fixes the dispersion of size distributions and the desired sharpening of size selection via manipulating the physical degrees of freedom at the cluster source should be approached with less pessimism.

## 6. Appendix

### Normal and Lognormal Distributions

The normal probability's cumulative is $C_{norm} := \frac{1}{2}\left[1 + \text{erf}\left(\tilde{n}/\sqrt{2}\right)\right]$. The normal distribution follows as $dC_{norm}/d\tilde{n} = \left(\sqrt{2\pi}\exp[\tilde{n}^2/2]\right)^{-1}$. The modal equals the mean $\text{MODAL}_m = \langle m \rangle$. The distribution of the *M* is according to Eq. (5) just the one for *m* but divided by *M* and thus leads to the LN distribution in Eq.(6). Most variables of interest, like the mean or modal, are easiest calculated by going back to the normal expressions. The expansion $M = e^m = \sum_{i=0}^{\infty}\left(m^i/i!\right)$ is necessary to express the p$^{th}$ moment

$$\langle M^p \rangle = e^{p\langle m \rangle + p^2 \Delta_m^2/2} \tag{11}$$

Hence, the mean <*M*> is larger than the modal $\text{MODAL}_M = e^{\langle m \rangle - \Delta_m^2}$ and transforms always with a shift $\langle M^p \rangle = \langle M \rangle^p e^{p(p-1)\Delta_m^2/2}$. For example, given a LN cluster size distribution with number expectation <*N*>, the expectation for the liquid drop model radius $R = r_S N^{1/3}$ becomes $\langle R \rangle = r_S \langle N \rangle^{1/3} e^{-\Delta_n^2/9}$ and is smaller than $r_S \langle N \rangle^{1/3}$. At this point one can straightforwardly derive Eq. (7) or equivalently $\Delta_m^2 = \ln[1 + 1/d_M]$. A standard deviation is generally the preferred measure of deviation; the famous "two sigma" is the standard width. However, this can become problematic for strongly skewed functions and those peaked close to the limit of their support. Here, the length $2\Delta_M$, when centered at the modal, likely reaches back below $M < 0$, i.e. $MODAL_M - \Delta_M$ may be negative. If $M$ cannot be negative, the FWHM is often employed instead. This popular measure of deviation is centered at Modal($M$) and extends laterally to where the PDF is only $X = ½$ of that maximum. This makes it difficult to relate it to standard deviations in case of the LN: The "full width at X maximum" is FW$X$M = $(E_+ - E_-)$, with $E_\pm = \exp[e_\pm]$ and $e_\pm = \langle m \rangle - \Delta_m^2 \pm \Delta_m \sqrt{-2\ln X}$. From the normal distribution's point of view, all this is unnecessary. $M < 0$ is excluded because no choice of measure stretches below $m = -\infty$. The general deviation FW$X$M is in $m$-space simply $(e_+ - e_-) = \Delta_m \sqrt{-8\ln X}$ (but not centered at <$m$>).

## Exponential and Gamma Distributions

The regularized gamma ($\Gamma$) function is the cumulative: $C_\Gamma = 1 - \Gamma(d_N, d_N Z)/\Gamma(d_N)$. $Z$ is the normalized variable $Z = N/\langle N \rangle$. From this, the $\Gamma$ distribution

$$\text{PDF}_\Gamma^{-1} = N e^{d_N Z} (d_N Z)^{-d_N} \Gamma_{(d_N)} \qquad (12)$$

is derived as shown in Section 2. For $d_N > 1$ there is a modal value and the $\Gamma$ distribution looks similar to a LN in that case. $d_N = 1$ implies the cumulative probability $C_{\text{EXP}} = C_{\Gamma(d_N=1)} = 1 - e^{-z}$. The distribution of $M$ follows as $dC_{\text{EXP}}/dM = e^{-Z}(dZ/dM)$ or

$dC_{\text{EXP}}/dM = \dfrac{Z}{M\Delta_m} e^{-Z}$. We yield $\text{PDF}_{\text{EXP}} = dC_{\text{EXP}}/dN = e^{-Z}/\langle N \rangle$ and this is again just

$n$'s distribution divided by $N$ [Eq. (5)], therefore $dC_{\text{EXP}}/dn = \langle N \rangle^{-1} e^{n - (e^n/\langle N \rangle)}$. This could

be called the "exp-exponential" distribution to be consistent with the usual "log" that is

added to "normal". While a monotonic exponential decline has no modal value

(maximum), in $n$-space the modal equals ln<$N$>. It holds furthermore $\langle N \rangle = e^{\langle n \rangle + \gamma}$

(similar to LN's $\langle N \rangle = e^{\langle n \rangle + \Delta_n^2/2}$), where $\gamma$ is the Euler-Mascheroni constant:

$\gamma = \lim\limits_{G \to \infty}\left[-\ln G + \sum_{g=1}^{G} 1/g\right] \approx 0.5772$. Thus, $dC_{\text{EXP}}/dn = e^{n - \langle n \rangle - \gamma - e^{n - \langle n \rangle - \gamma}}$ and the standard

deviation is derived to be $\Delta_n = \pi/\sqrt{6} \approx 1.28$.